\documentstyle[12pt,epsfig]{article}
\begin{document}
\newcommand {\klref}[1]{(\ref{#1})}
\newcommand {\nucleus}[2] {\mbox{${}^{#2}{#1}$}}
   \newcommand {\He} {\nucleus{\rm He}{4}}
   \newcommand {\Be} {\nucleus{\rm Be}{8}}
   \newcommand {\Bxi} {\nucleus{\rm B}{11}}
   \newcommand {\Cxi} {\nucleus{\rm C}{11}}
   \newcommand {\Bxii} {\nucleus{\rm B}{12}}
   \newcommand {\Cxii} {\nucleus{\rm C}{12}}
   \newcommand {\Nxii} {\nucleus{\rm N}{12}}
   \newcommand {\Oxvi} {\nucleus{\rm O}{16}}
   \newcommand {\Fxvi} {\nucleus{\rm F}{16}}
   \newcommand{\CCmCxiiX}{\Cxii{}($\overline{\nu}_\mu,\mu^+$)X}
   \newcommand{\CDmCxiiX}{\Cxii{}($\nu_{\mu},\mu^-$)X}
   \newcommand{\CCmCxiig}{\Cxii{}($\overline{\nu}_\mu,\mu^+$)\Bxii{}$_{g.s.}$}
   \newcommand{\CDmCxiig}{\Cxii{}($\nu_{\mu},\mu^-$)\Nxii{}$_{g.s.}$}
   \newcommand{\NNCxiip} {\Cxii{}($\nu, \nu^{\prime}$p)\Bxi{}}
   \newcommand{\NACxiip} {\Cxii{}($\overline{\nu},
                                   \overline{\nu}^{\prime}$p)\Bxi{}}
   \newcommand{\NNCxiin} {\Cxii{}($\nu, \nu^{\prime}$n)\Cxi{}}
   \newcommand{\NACxiin} {\Cxii{}($\overline{\nu},
                                   \overline{\nu}^{\prime}$n)\Cxi{}}

   \newcommand{\brcite}[1]{[\cite{#1}]}
%
%
\baselineskip18pt
\begin{titlepage}
\title{
Strangeness in the nucleon and the ratio of proton-to-neutron
neutrino-induced quasi--elastic yield
}
\author{
E.~Kolbe \\
{\normalsize Institut f\"ur Physik der Universit\"at Basel} \\
{\normalsize Klingelbergstrasse 82} \\
{\normalsize CH--4056 Basel, Switzerland} \\
\\
S.~Krewald \\
{\normalsize Institut f\"ur Kernphysik} \\
{\normalsize Forschungsanlage J\"ulich} \\
{\normalsize D--52425 J\"ulich, Germany} \\
\\
and \\
\\
H.~Weigel \\
{\normalsize Institut f\"ur Theoretische Physik} \\
{\normalsize Universit\"at T\"ubingen} \\
{\normalsize D--72076 T\"ubingen, Germany} \\
}
\date{December 1996}
\end{titlepage}
\maketitle
\vskip2cm
\begin{abstract}
The electroweak form factors of the nucleon as obtained within a 
three flavor pseudoscalar vector meson soliton model are employed to 
predict the ratio of the proton and neutron yields from \Cxii{}, which 
are induced by quasi--elastic neutrino reactions. These predictions 
are found to vary only moderately in the parameter space allowed by 
the model. The antineutrino flux of the up--coming experiment determining 
this ratio was previously overestimated. The corresponding correction is 
shown to have only a small effect on the predicted ratio. However, 
it is found that the experimental result for the ratio crucially
depends on an accurate measurement of the energy of the knocked 
out nucleon.
\end{abstract}
\vskip1cm
%
\vfil\eject
\pagestyle{plain}
\leftline{\large \bf 1. Introduction}
\bigskip

By means of a continuum random phase approximation (RPA) 
\cite{Bu91,Ko92} it has been shown \cite{Ga92,Ga93} that the ratio 
$R_y$ of the proton-to-neutron neutrino-induced quasi--elastic yields 
from an isoscalar nucleus provides a relation between the nucleon 
matrix elements of the strange quark vector and axial vector 
currents $\langle N|{\bar s}\gamma_\mu s|N\rangle$ and
$\langle N|{\bar s}\gamma_\mu\gamma_5 s|N\rangle$, respectively.
This is especially interesting because it allows one to determine 
the singlet matrix element $g_A^{(0)}$ of the axial vector current 
(the so--called proton spin puzzle \cite{Br88}) without assuming flavor 
$SU(3)$ symmetry. This symmetry is commonly employed to relate matrix 
elements of currents which conserve the flavor charges to those which 
change this quantum number. The latter are measured in semi--leptonic 
hyperon decays.

The purpose of the present paper is three--fold. Firstly, we will 
update the earlier calculations \cite{Ga92,Ga93} by substituting more 
accurate energy spectra for the LAMPF neutrino and antineutrino beam 
which recently became available \cite{BuPC}. This spectrum enters the 
computation of the quasi elastic yields by folding it into the 
cross--section for the (anti) neutrino -- nucleus reaction \cite{Ko92}.
In the second part we will adopt the $SU(3)$ Skyrme model with vector 
mesons \cite{Ka84} to compute the relevant momentum dependent form 
factors of the nucleon. This makes possible a prediction for $R_y$ 
because these form factors serve as the essential ingredients for 
the nuclear model calculation developed in refs. \cite{Ga92,Ga93}. 
Although the incorporation of vector meson fields into the $SU(3)$ 
Skyrme model represents a severe complication \cite{Pa91} these fields 
are unavoidable for a realistic description of static baryon properties. 
In particular these fields account for short range effects, which are 
particularly important to gain a realistic, non--zero value for 
$g_A^{(0)}$ \cite{Jo90}. The simple Skyrme model of pseudoscalars 
predicts $g_A(q^2)\equiv0$ \cite{Pa89}. Such short range 
components are also necessary to gain a non--zero result for the 
neutron--proton mass difference \cite{Ja89}. On the basis that 
this model reproduces the experimental two--flavor form factors only 
within an error of 20--30\% (which already represents an improvement 
compared with the pure $SU(3)$ Skyrme model) one might want to argue 
that other approaches \cite{Mu94} may be more appropriate to compute 
strange form factors. However, this model is unique in the sense that 
it treats vector and axial vector currents simultaneously and contains 
a consistent and systematic treatment of symmetry breaking effects. The 
resulting predictions for $R_y$ turn out to be quite insensitive to 
the model parameters. This indicates that the strange form factors may 
independently be considered from those in the two flavor subspace. A
recent study \cite{Ba96} has shown that the ratio $R_y$ is not very 
sensitive to the nuclear model but significantly depends on the 
prediction for $\langle N|{\bar s}\gamma_\mu\gamma_5 s|N\rangle$. In 
this respect a precise measurement of $R_y$ by the LSND-Collaboration 
at LAMPF \cite{LSND} will be discriminating on the nucleon model.
In section 3 we will briefly review the computation of 
the relevant form factors in this model after having outlined (in 
section 2) their relevance for the (anti)neutrino scattering off nuclei 
in the context of the continuum RPA. 

Finally we will point out a possible source of uncertainty in the 
experiment associated with the sensitive dependence of $R_y$ on the 
energy thresholds for the integrated proton and neutron yields.
The corresponding results together with a comprehensive discussion 
will be presented in section 4.

\bigskip
\leftline{\large \bf  2. The $\mbox{\boldmath $\nu$}$ --
$\bar{\mbox{\boldmath $\nu$}}$ nuclei reaction}
\bigskip
At low energies (compared to the electro--weak scale) the interaction
of neutrinos with matter is described by the current--current 
interaction introducing the Fermi coupling constant $G_F$
\begin{eqnarray}
{\cal L}_{\rm eff}=-\frac{G_F}{\sqrt2}
\left(j_\mu^{(+)}J^{\mu(-)}+j_\mu^{(0)}J^{\mu(0)} +{\rm h.\ c.}\right).
\label{leff}
\end{eqnarray}
Here $j_\mu^{(a)}$ denotes the leptonic part of the electro--weak 
current\footnote{This shorthand notation is assumed to include 
the dependence on the Weinberg angle, $\Theta_{\rm W}$.} 
while $J^{\mu(a)}$ represents the hadronic counterpart. 
Of course, these currents break parity in a maximal way, {\it i.e.}
$j_\mu^{(a)}=v_\mu^{(a)}-a_\mu^{(a)}$ with $v$ and $a$ referring to 
the vector and axial--vector currents. The hadronic (axial)vector 
currents ($V$ and $A$) are defined similarly.

To describe neutrino scattering off nuclei in a perturbative 
treatment of ${\cal L}_{\rm eff}$ we will employ the continuum RPA
approach (as a mean field model) of ref.\cite{Bu91,Ko92} in order 
to compute matrix elements of $J^{\mu(a)}$ when sandwiched between 
nuclei states\footnote{The expression for the scattering cross 
section is given in eq (11) of ref. \cite{Ko92}.}. In this nuclear 
model the interaction between the constituents of the nucleus is 
described by combining the usual RPA treatment with a correct 
description of the particle states in the continuum, {\it i.e.} 
the excited many-body states are coherent superpositions of 
one-particle-one-hole ($1p$-$1h$) excitations obeying the proper 
Coulomb boundary conditions for scattering states. For details 
on this approach we refer to the literature \cite{Bu91,Ko92}. Here 
we restrict ourselves to a brief description of its basic 
features and ingredients. The continuum RPA provides a good description 
of the nuclear ground state while the excited states are generic 
continuum states possessing a $1p$-$1h$ structure. Final state 
interactions are accounted for by a realistic (finite range) residual 
interaction derived from the Bonn meson exchange potential 
\cite{Na87,Ma87}. It should also be mentioned that this treatment 
assumes that the reactions proceed predominantly via the knocked 
out nucleon. As the main result of the continuum RPA approach 
the scattering cross section of the (anti)neutrino nuclei reaction 
is expressed in terms of the nucleon matrix elements \begin{eqnarray}
\langle N | J^{\mu(a)} | N \rangle\ , \quad a=0,...,8\ ,
\label{nuclmat}
\end{eqnarray}
where $| N \rangle$ represents a nucleon isospinor. Since the 
nucleon is an extended object form factors for this matrix elements 
are mandatory. The Lorentz covariant expressions for the 
(axial)vector current matrix elements\footnote{We omit the 
induced form factor of the axial current.} of the flavor conserving 
components ($a=0,3,8$) read
\begin{eqnarray}
\langle N^\prime | V^a_\mu |N \rangle &=&
\bar{u}_N(p^\prime)\left[F_1^a\left(q^2\right)\gamma_\mu
+F_2^a\left(q^2\right)i\sigma_{\mu\nu}\frac{q^\nu}{2M}\right]u_N(p)
\nonumber \\
\langle N^\prime | A^a_\mu |N \rangle &=&
\bar{u}_N(p^\prime)\left[G^a_A\left(q^2\right)
\gamma_\mu\gamma_5\right]u_N(p)
\label{defff}
\end{eqnarray}
where $u_N(p)$ denotes the proton Dirac spinor. The neutron form 
factors are obtained from isospin covariance. The linear combinations 
relevant for the processes under consideration are 
\cite{Ga92,Ga93}
\begin{eqnarray}
F_i(q^2)&=&
\pm\left(\frac{1}{2}-{\rm sin}^2\Theta_{\rm W}\right)F_i^3(q^2)
-{\rm sin}^2\Theta_{\rm W}F_i^{I=0}(q^2)-\frac{1}{2}F_i^s(q^2)
\quad (i=1,2),
\nonumber \\
G(q^2)&=&\mp\frac{1}{2}G_A^3(q^2)+\frac{1}{2}G_A^s(q^2),
\label{relff}
\end{eqnarray}
where $I=0$ and $s$ denote the non--strange and strange combinations 
of $a=0,8$, respectively. The upper (lower) sign refers to the 
proton (neutron). The vector form factors $F_i$ are more conveniently 
expressed in terms of electric and magnetic combinations
\begin{eqnarray}
G^a_E(q^2)=F^a_1(q^2)-\frac{q^2}{4M^2}F^a_2(q^2) \quad {\rm and} \quad
G^a_M(q^2)=F^a_1(q^2)+F^a_2(q^2).
\label{gegm}
\end{eqnarray}
The charge matrix is given by 
${\cal Q}=(\lambda^3+\lambda^8/\sqrt3)/2$ hence one may adopt the 
physically more transparent basis consisting of the electro--magnetic 
form factors of the nucleon, $G_{E,M}^{p,n}(q^2)$. Due to isospin 
covariance $G^3_A(q^2)$ is identical to the axial charge $G_A(q^2)$
of the nucleon extracted from neutron $\beta$--decay. The computation 
of these form factors in the Skyrme model with vector mesons will 
be reviewed in the next section.

\bigskip

\leftline{\large \bf  3. Nucleon form factors in the vector 
meson Skyrme model} 
\bigskip

Our starting point is a three--flavor chirally invariant 
theory for pseudoscalar and vector mesons. The model Lagrangian 
also contains abnormal parity terms \cite{Ka84} to accommodate 
processes like $\omega\rightarrow3\pi$. A minimal set of 
symmetry breaking terms, which transform as 
$\left({\bar 3}\times3+{\rm h.c.}\right)$ under 
the chiral group $SU_L(3)\times SU_R(3)$, is included \cite{Ja89} to 
account for different masses and decay constants. This effective 
theory contains topologically non--trivial static solutions, which 
are constructed by imposing {\it ans\"atze} in the isospin subgroup
\begin{eqnarray}
\xi_\pi(\mbox{\boldmath $r$})={\rm exp}\left(\frac{i}{2}
\hat{\mbox{\boldmath $r$}}\cdot\mbox{\boldmath $\tau$}F(r)\right),
\quad
\omega_0(\mbox{\boldmath $r$})=\omega(r)
\quad {\rm and} \quad
\rho_{i,a}(\mbox{\boldmath $r$})=
\frac{G(r)}{r}\epsilon_{ija}\hat r_j\ ,
\label{solan}
\end{eqnarray}
while all other field components vanish. Here 
$\xi_\pi={\rm exp}\left(i\mbox{\boldmath $\pi$}\cdot
\mbox{\boldmath $\tau$}/2f_\pi\right)$ refers to the non--linear 
realization of the pion fields. The radial functions are determined 
by extremizing the static energy functional subject to boundary 
conditions appropriate to the topological sector of winding number 
one. Motivated by the large $N_C$ studies of QCD these solitons are 
identified as the baryons with unit baryon number \cite{Wi79}. 
Unfortunately the field configuration (\ref{solan}) carries neither 
good spin nor flavor quantum numbers, hence an appropriate projection 
has to be performed. Also strange flavors are not yet excited. These 
two features are accounted for by introducing time dependent 
collective coordinates $A(t)$ for the (approximate) zero modes 
associated with the $SU(3)$ vector rotations:
$\xi(\mbox{\boldmath $r$},t)=
A(t)\xi_\pi(\mbox{\boldmath $r$})A^{\dag}(t)$
and similarly for the vector meson nonet. Furthermore field 
components, which vanish classically like {\it e.g.} the kaons, 
are induced by the collective rotations. These induced fields 
turn out to be proportional to the angular velocities
\begin{eqnarray}
A^{\dag}\dot A=\frac{i}{2}\sum_{a=1}^8\lambda^a\Omega^a
\label{defvel}
\end{eqnarray}
and exist for both, rotations into strange ($a=4,..,7$) and 
non--strange ($a=1,2,3$) directions. In order to compute the induced 
fields {\it ans\"atze} are chosen, which are consistent with the 
parity as well as Lorentz and isospin structure of the considered 
mesons. Altogether eight real radial functions are introduced, which 
solve inhomogeneous linear differential equations derived from a 
variational principle to the moments of inertia \cite{Pa91}. The 
classical fields (\ref{solan}) act as source terms, which stem 
from the abnormal parity parts of the effective meson theory. The 
collective Lagrangian $L=L(A,\{\Omega^a\})$ is extracted and the 
collective coordinates are quantized canonically \cite{Ad83}
\begin{eqnarray}
R_a=-\frac{\partial L(A,\{\Omega^b\})}{\partial \Omega^a}
\label{quant}
\end{eqnarray}
providing a linear relation between the angular velocities $\Omega^a$
and the right generators of $SU(3)$, $R_a$. The fact that these 
objects are operators is reflected by the commutation 
relations\footnote{The right generators of $SU(3)$ are identified 
in the body fixed system of the rotating soliton.}
$[R_a,R_b]=-f_{abc}R_c$, with $f_{abc}$ being the structure 
constants of $SU(3)$. The resulting collective Hamiltonian may be 
diagonalized exactly \cite{Ya88} yielding the spectrum of the 
low--lying $\frac{1}{2}^+$ and $\frac{3}{2}^+$ baryons. In the 
absence of flavor symmetry breaking the eigenfunctions $\Psi(A)$ of 
this Hamiltonian are $SU(3)$ D--functions associated with a certain 
representation ({\it e.g.} the $\mbox{\boldmath $8$}$ for the 
nucleon). Once flavor symmetry breaking is included these 
D--functions become distorted reflecting the admixture of higher 
dimensional representations as for example the 
$\overline{\mbox{\boldmath $10$}}$ or $\mbox{\boldmath $27$}$.

Extending the action to account for electro--weak interactions allows 
one to derive covariant expressions for the (axial--)vector currents 
from the terms linear in the corresponding gauge fields. Substituting 
the above described {\it ans\"atze} and eliminating the angular
velocities in favor of the generators (\ref{quant}) leaves the 
currents as linear combinations of radial functions and operators in 
the space of the collective coordinates $A$. The former are given in 
terms of the classical and induced profile functions of the meson 
fields. The matrix elements of the currents, which eventually provide 
the relevant form factors, are computed in two steps. Firstly, the 
radial functions are Fourier--transformed yielding the momentum 
dependence of the form factors in the Breit frame \cite{Br86}. 
Secondly, an $SU(3)$ ``Euler--angle" representation for $A(t)$ is 
employed \cite{Ya88} to parametrize the $SU(3)$ operators as well as 
the exact eigenfunctions $\Psi(A)$. This makes possible the evaluation 
of the spin and flavor parts in the matrix elements of the currents.

In ref. \cite{Pa91} the parameters of the model Lagrangian, which could 
not be determined from the meson sector, were adjusted to provide a 
best fit to the baryon mass differences. This has lead to reasonable, 
though not perfect, agreement for the electro--magnetic and axial 
form factors. The strange form factors, which are obtained by
considering different flavor components of the currents, were predicted.
In table \ref{ta_ff} we display the results for zero momentum 
transfer together with results corresponding to parameter sets which 
improve on the axial charge of the nucleon and/or the magnetic 
magnetic moment of the nucleon. Also displayed are the predicted 
strange form factors at zero momentum transfer. Generally we find that 
all parametrizations predict the strangeness form factors to be 
negative and small in magnitude.
\begin{table}
\caption{\label{ta_ff}Predicted form factors at zero momentum 
transfer. Three sets of parameters in the model Lagrangian 
are used. ($\times$): best fit to the baryon mass differences,
(*): reproducing the experimental value for $G_A(0)$,
(+): best fit to $G_A(0), G_M^p(0)$ and $G_M^n(0)$. Furthermore 
the results associated with $SU(3)$ symmetric wave--functions 
are presented for the parameter set ($\times$).}
~
\newline
\centerline{
\begin{tabular}{| c | c c c c c |}
\hline
& $G_A(0)$ & $G_M^p(0)$ & $G_M^n(0)$ & $G_A^s(0)$ & $G_M^s(0)$ \\
\hline
$\times$& 0.94 & 2.35 & -1.86 & -0.030 & -0.055 \\
$*$& 1.25 & 3.23 & -2.86 & -0.015 & -0.035 \\
$+$& 1.02 & 2.57 & -2.11 & -0.028 & -0.050 \\
\hline
$SU(3)$ sym. & 0.88 & 2.48 & -1.54 & -0.058 & -0.559 \\
\hline
Exp. & 1.25 & 2.79 & -1.91 & --- & --- \\
\hline
\end{tabular}
}
\end{table}
In fig. \ref{fi_one} we display the predicted momentum dependence of 
the strange vector form factor $G_S(q^2)$. It is interesting to note 
that this quantity exhibits a local minimum at 
$q^2\approx0.2{\rm GeV^2}$ before dropping to zero at larger $q^2$. 
Apparently the frequently employed multipole parametrizations will not 
accurately reproduce this feature of $G_S(q^2)$. 
Kinematical corrections have not been 
incorporated here because they are not relevant for the momentum 
transfers of the present problem \cite{Ji91}. When $SU(3)$ symmetric 
wave--functions are employed to compute the spin and flavor parts of 
the matrix elements the strange magnetic moment increases by one order 
of magnitude as compared to the use of the exact eigenfunctions. This 
is not surprising since the use of these wave--functions assumes that 
strange quarks are as light as the non--strange ones. Obviously 
virtual strange degrees of freedom are easily excitable in that case. 

The realistic amount of flavor symmetry breaking yields a small and 
negative value for the strange magnetic, even smaller than other 
estimates do \cite{Mu94,Jaf89,Ja90,Ha95}. To some extend all these 
predictions seem to be contradictory to the preliminary results from 
the experiment at MIT--Bates: 
$G_M^s(0)=0.46\pm0.36\pm0.08\pm0.18$. The first erros is statistical,
the second due to the background and the third takes account of 
uncertainties in the radiative corrections entering the analysis
\cite{Sch96}.  This measurement apparently favors a 
sizable positive value, however, due to large errors small negative 
values are not ruled out either. On the other hand, a 
large negative strange magnetic moment, as some calculations predict 
\cite{Le95,Kim96}, is unlikely. Estimates of the strange axial charge 
$G^s_A(0)$ are closely connected to the issue of the proton spin 
puzzle \cite{Br88}. Employing flavor symmetry and using data from 
semi--leptonic hyperon decays results in $G^s_A(0)\approx -0.1$. 
When deviations from flavor symmetry are taken into account this 
number may easily be reduced and even a zero value is possible 
\cite{Jo90,Li95}.

\bigskip
\leftline{\large \bf 4. Results and discussion}
\bigskip

Before employing the above computed form factors to evaluate $R_y$ 
we will (in fig.~\ref{fi_two}) compare the old energy spectrum for the 
LAMPF neutrino beam with the new and more accurate one \cite{BuPC}. As 
can be observed when comparing the solid and dashed lines the energy 
distribution of the muon--neutrinos have not changed significantly. 
However, the new muon--antineutrino flux is different in shape and 
magnitude from the old one. Previously the latter was estimated by 
simply assuming it contributed 20\% to the total ({\it i.e.} 
$\tilde \nu= \nu + \overline\nu$) neutrino beam. The improved 
spectrum for the antineutrinos is peaking at lower energies hence the 
yield induced by the antineutrinos is significantly decreased. The
reason is that the extraction of $R_y$ requires a threshold of 
$E_N>60$~MeV for the energy of the knocked out nucleon to discriminate 
against events from elastic neutrino scattering on free protons.

We have next substituted the data of the improved spectra to predict 
$R_y$ as a function of the strange quark axial charge, $G_A^s(0)$
for different values of the strange quark magnetic moment $G_M^s(0)$. 
A dipole parametrization, which may not be very realistic {\it cf.}
fig \ref{fi_one}, has been employed for the momentum dependence 
\cite{Ga92}. In fig.~\ref{fi_thr} the results are displayed for both 
the quasi--elastic antineutrino-- and neu\-tri\-no-in\-duced reactions
on \Cxii{}. The range assumed for $G_M^s(0)$ is taken from the 
estimate of refs. \cite{Jaf89,Ja90}\footnote{More recently a similar
analysis which attempts to make contact with perturbative QCD has been 
performed in ref \cite{Ha95}. These authors consider their result 
$G_M^s(0)=-0.24\pm0.03$ as an upper bound (in magnitude). However, 
neither of these calculations incorporates the feature of flavor 
symmetry breaking.}. For the non--strange form factor also a dipole 
approximation has been adopted. As the reduction factor is the same 
for the proton and the neutron, the ratio 
$\sigma(\overline\nu,\overline\nu^\prime p)/
 \sigma(\overline\nu,\overline\nu^\prime n)$ 
is not effected when using the improved spectra. Hence the almost 
linear dependence on $G_A^s(0)$, which was previously found ({\it cf.}
fig.~4 in~\cite{Ga93}), is recovered, although the absolute 
antineutrino induced cross sections have decreased by a factor of 
approximately 4.3. Since the interference terms contribute with 
opposite signs to the $\overline\nu$- and $\nu$-induced reactions, 
$R_y$ exhibits a stronger increase with $G_A^s(0)$ for the 
$\overline\nu$ induced reaction than for the one associated with 
the neutrino.

Fig.\ref{fi_thr} also contains the prediction for $R_y$ obtained 
from the momentum dependent form factors predicted by the 
$SU(3)$ Skyrme model with vector mesons as outlined in the 
preceding section.  While, for the case of antineutrino-induced 
reactions, the ratios are slightly different for the three parameters 
sets discussed above, they are practically indistinguishable for 
neutrino-induced scattering. As a consequence of the quite 
small value for $G_M^s(0)$ the prediction of this model is 
expected to be approximately given by the dashed lines in 
fig.~\ref{fi_thr}. However, this is not quite the case since the 
model does not exactly reproduce the non--strange form 
factors, {\it cf.} table \ref{ta_ff}.

In fig.~\ref{fi_four} we finally display $R_y(\tilde \nu=\nu+\bar\nu)$ as 
obtained when both the neutrino and antineutrino yields are included.
As a consequence of using the improved antineutrino spectrum, 
$R_y(\tilde \nu)$ is not quite as sensitive to $G_A^s(0)$ as 
previously \cite{Ga92,Ga93} estimated. On the other hand the 
sensitivity on $G_M^s(0)$ is increased. Furthermore the negligibly 
small $\overline\nu$-flux causes the curves of fig.~\ref{fi_two}b 
and fig.~\ref{fi_four} to be identical within 1\%.
From fig.~\ref{fi_four} we deduce $1.1\le R_y(\tilde \nu)\le1.2$
as the prediction obtained from the Skyrme model with vector mesons.
Adopting flavor symmetric wave--functions ({\it i.e.} $SU(3)$
D--functions) significantly increases this prediction,
$R_y(\tilde \nu)\approx1.4$. Thus an experimental determination 
of $R_y(\tilde \nu)$ will in particular serve as a test
of the quantization procedure proposed in ref. \cite{Ya88}. In this 
context it should be stressed that a precise measurement of $R_y$ will 
especially be suited to (dis)approve models for the baryons. The 
reason is that for this quantity the uncertainties inherited from 
the nuclear model will cancel almost completely. Barbaro {\it et al.} 
\cite{Ba96} have recently addressed this issue by comparing the single 
proton and neutron yields as well as their ratio $R_y$ in two extreme 
nuclear models with $G_A^s(0)$ being a parameter. It has then been 
observed that $R_y$ is almost independent of the nuclear model but very 
sensitive to $G_A^s(0)$, which, of course, is a prediction of the
nucleon model. On the other hand the separate yields may significantly 
vary when $G_A^s(0)$ is kept fixed but different nuclear models are 
employed.

In our treatment strange degrees of freedom enter solely by virtual
excitations of strange quark--antiquark pairs inside the nucleon. 
Let us also briefly coment on uncertainties which enter due to 
other effects.  One may  {\it e.g.} wonder whether or not the coupling 
of the leptonic current to the meson exchange currents between the 
constituents of the nucleus lead to additional contributions. This 
problem has been addressed by Musolf et al. \cite{Mu94a} for the case 
of \He. Although this nucleus is less complex than \Cxii\ it represents 
a $J=0$ and $T=0$ nucleus as well. The exchange terms were observed 
to considerably improve on the agreement of the non--strange isoscalar 
form factor of \He\ with experiment (see fig.~2 of ref. \cite{Mu94a}).
Nevertheless they have turned out not to be relevant 
for momentum transfers less than approximately $500{\rm MeV}$ for the 
strange vector form factor (see fig.~3 of ref. \cite{Mu94a}). From the 
energy spectrum displayed in fig.~\ref{fi_two} we hence conclude that 
our results will not significantly be modified by the contributions 
associated with the meson exchange currents. From a conceptual point 
the meson exchange terms are incorporated to satisfy current 
conservation for the many--body system as this requirement is not 
necessarily satisfied by the one--body current operator. One may 
estimate the corresponding uncertainties by employing various 
expressions for the current operator which are related by the 
continuity equation \cite{Fr85}. For different excitations of 
\Cxii\ Friar and Haxton \cite{Fr85} have compared the electric form 
factors predicted by various forms of the current operator with 
experimental data. They find that the form of the current operator 
which incorporates current conservation constraints only at zero 
momentum transfer ($q=0$) but exhibits the correct high $q$ behavior 
contains at least some meson exchange effects. It is exactly this 
form which enters our calculation. The feature of correct high $q$ 
behavior is shared by neither  the form of the current operator which 
is completely constrained by current conservation nor by the one 
which is not constrained at all.

We would also like to point out a possible source of uncertainty which 
is related to the measurement of the energy $E_N$ of the knocked out 
nucleon, which may be relevant for the interpretation of the
experiments. Comparing the ratios $R_y$ in fig.~\ref{fi_four} with the 
previous results (fig.~5 in ref.~\cite{Ga93}), we notice that the curves 
differ by a constant shift of approximately 15\% in $R_y(\tilde{\nu})$. 
While for vanishing strangeness contributions to the form factors 
the present calculation (dashed line at $G_A^s(0)=0$ in 
fig.~\ref{fi_four}) gives the consistent value $R_y=1.0$, we extract
$R_y \approx 0.85$ from fig.~5 in ref.~\cite{Ga93}. 
This shift is caused by adopting different thresholds for the integrated
neutron-yield. In ref.~\cite{Ga93} $d\sigma/dE_N$ has been integrated 
over the energy of the emitted nucleon starting at  
$E_N=E_p=E_n+2.77$~MeV $>60$~MeV ({\it i.e.} as a consequence of 
the Coulomb--shift the energy threshold of the neutrons was taken to 
be $E_n^0=57.23$~MeV), contrary to the results shown in 
fig.~\ref{fi_two} and fig.~\ref{fi_four}, where we have set the 
thresholds to $E_N^0=E_p^0=E_n^0=60$~MeV. This crucial dependence of 
$R_y$ on the $E_p$-- and $E_n$--thresholds is linked to the 
steep decrease of the neutrino flux (fig.~\ref{fi_two}) 
between $70$ and $200$~MeV, which also causes the differential cross 
section $d\sigma/dE_N$ to strongly decrease with increasing $E_N$.
We investigated this effect in more detail by slightly varying 
$E_N^0$ for protons and neutrons. The results for neutrino-induced
reactions are summarized in table \ref{ta_pnr} (where we have set
$G_A^s=G_M^s=0$, for simplicity).
\begin{table}
\caption{\label{ta_pnr}The ratios $R_y$ for neutrino-induced nucleon 
knockout on \Cxii{} (and for $G_A^s(0)=G_M^s(0)=0$) as a function 
of different thresholds for protons ($E_p^0$) and  neutrons ($E_n^0$).}
\begin{center}
\begin{tabular}{|c|c|r|} \hline 
   $E_p^0$ [MeV] & $E_n^0$ [MeV] & $R_y(\nu)$ \\ \hline 
      60.0 & 60.0   & 1.004 \\
      60.0 & 59.0   & 0.943 \\
      60.0 & 61.0   & 1.069 \\
      61.0 & 60.0   & 0.943 \\
      61.0 & 59.0   & 0.885 \\
      61.0 & 58.0   & 0.832 \\
      61.0 & 61.0   & 1.004 \\ \hline
\end{tabular}
\end{center}
\end{table}
It is apparent that the experimentally determined value for 
$R_y$ crucially depends on an accurate measurement of the energy 
$E_N$ of the knocked out nucleon. A relative shift of only $1$~MeV 
between the thresholds $E_p^0$ and $E_n^0$ will cause an error 
for $R_y$ of $\approx 6$\%. We find the same $\approx 6$\% error for 
non--zero $G_A^s$ and $G_M^s$ as well. It should, however, be remarked 
that the absolute value of the $E_N$--threshold has no influence, as 
long as $E_p^0$ and $E_n^0$ are identical. This independence was 
already pointed out previously, see fig.~3 of Ref.~\cite{Ga93}.

\vskip1.5cm
\leftline{\large\it Acknowledgements}
\bigskip

One of us (HW) acknowledges support by the Deutsche 
Forsch\-ungsgemeinschaft (DFG) under contracts We1254/2--1 and
Re856/2--2. 

\newpage

\vskip1.5cm

\pagestyle{empty}
\newpage
\section*{Figure captions}
\begin{figure}[h]
\caption{\label{fi_one}The momentum dependent strange vector 
magnetic form factor $G_M^s$ in the $SU(3)$ vector meson Skyrme 
model (left panel). For completeness the strange axial vector form 
factor $G_A^s$ is also given (right panel). Different parameter sets 
have been used, {\it cf.} table \protect\ref{ta_ff}.}
\vskip1cm
\caption{\label{fi_two}Comparison of the previous and updated 
(anti)neutrino spectra for the LAMPF decay-in-flight neutrino 
source. The solid, dotted and dashed lines refer to the updated 
neutrino, updated antineutrino and previous (anti) neutrino spectra,
respectively. The antineutrino spectra are always scaled by a factor 
four.}
\vskip1cm
\caption{\label{fi_thr}Ratio of integrated proton-to-neutron yield 
for quasi--elastic antineutrino- (upper part) and neutrino-induced 
(lower part) reactions on ${}^{12}$C as a function of $-G_A^s(0)$ 
for different values of $G_M^s(0)$ within the theoretically estimated 
regime \protect\cite{Jaf89,Ja90}. The symbols indicate the predictions 
of the $SU(3)$ Skyrme model with vector mesons, {\it cf.} 
table~\protect\ref{ta_ff}. Their location on the horizontal axis 
reflects the associated prediction for $G_A^s(0)$.}
\vskip1cm
\caption{\label{fi_four}Same as fig. \protect\ref{fi_thr}, for the 
sum of antineutrino- and neutrino-induced yield.}
\end{figure}

\newpage
\begin{figure}[htb]
\centerline{\hskip 0.0cm 
\epsfig{figure=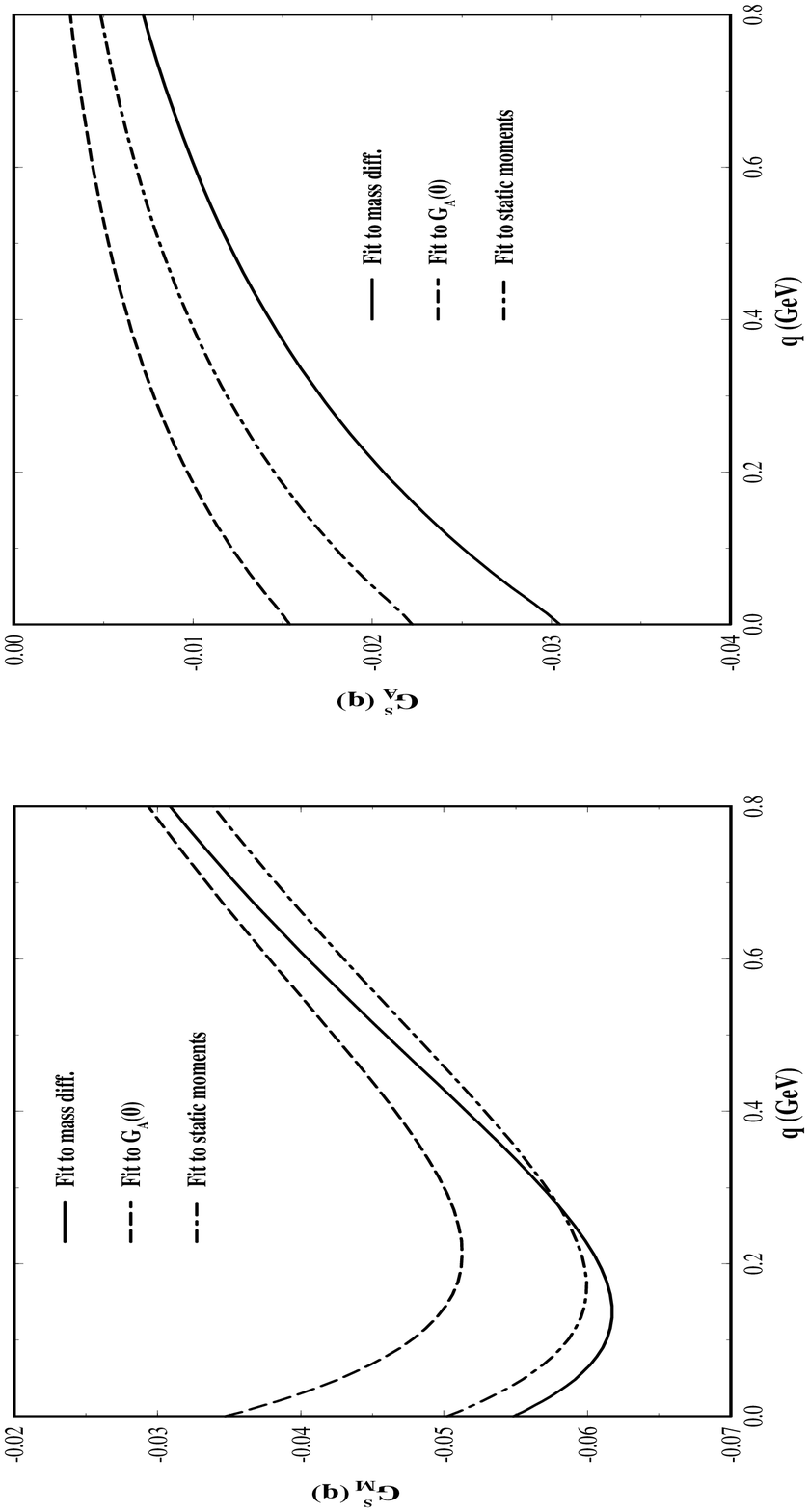,height=22.5cm,width=16.0cm}}
\end{figure}
\begin{figure}[htb]
\centerline{\hskip -1.5cm
\epsfig{figure=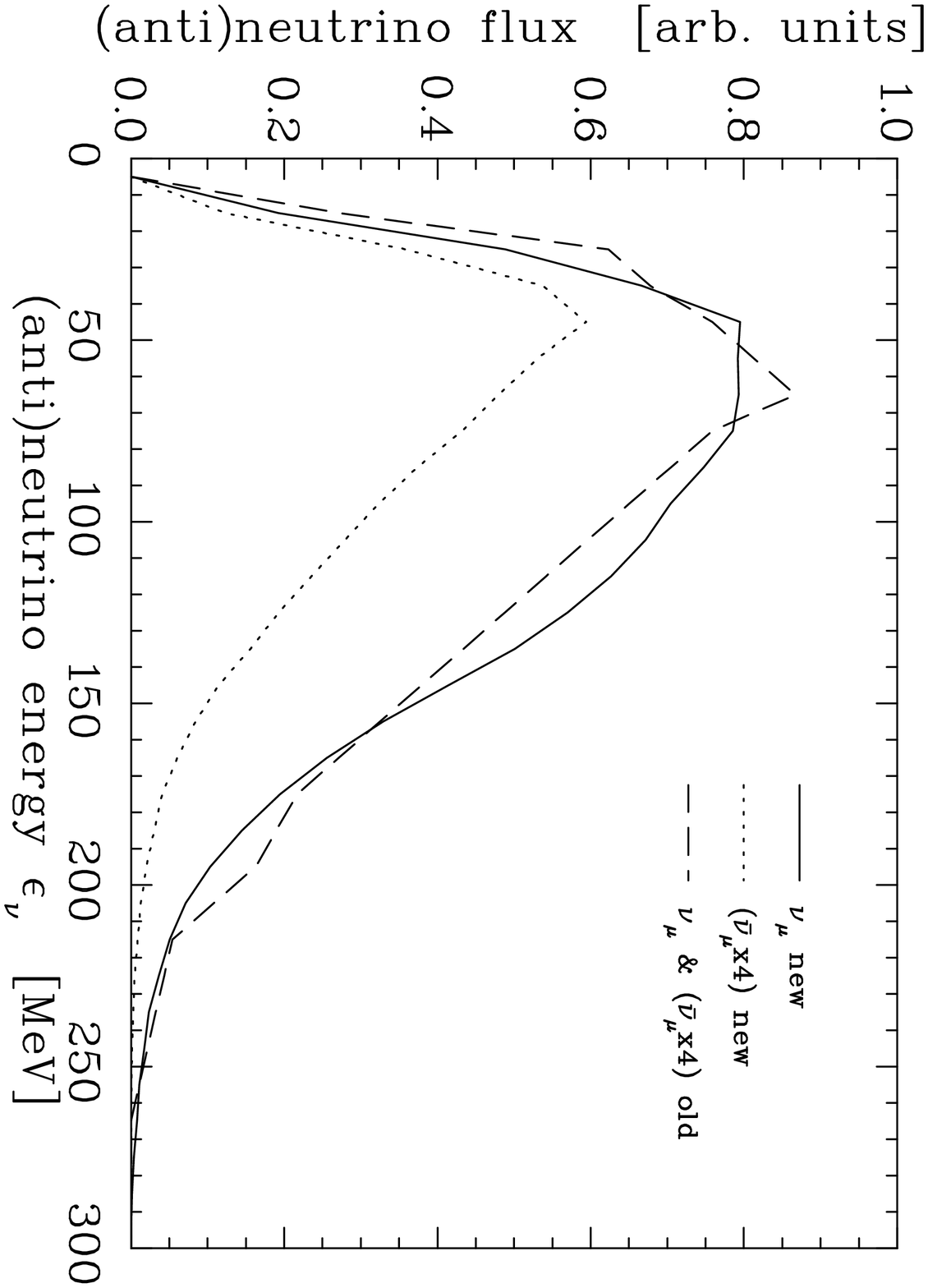,height=20.0cm,width=16.0cm}}
\end{figure}
\begin{figure}[htb]
\centerline{\hskip -1.5cm
\epsfig{figure=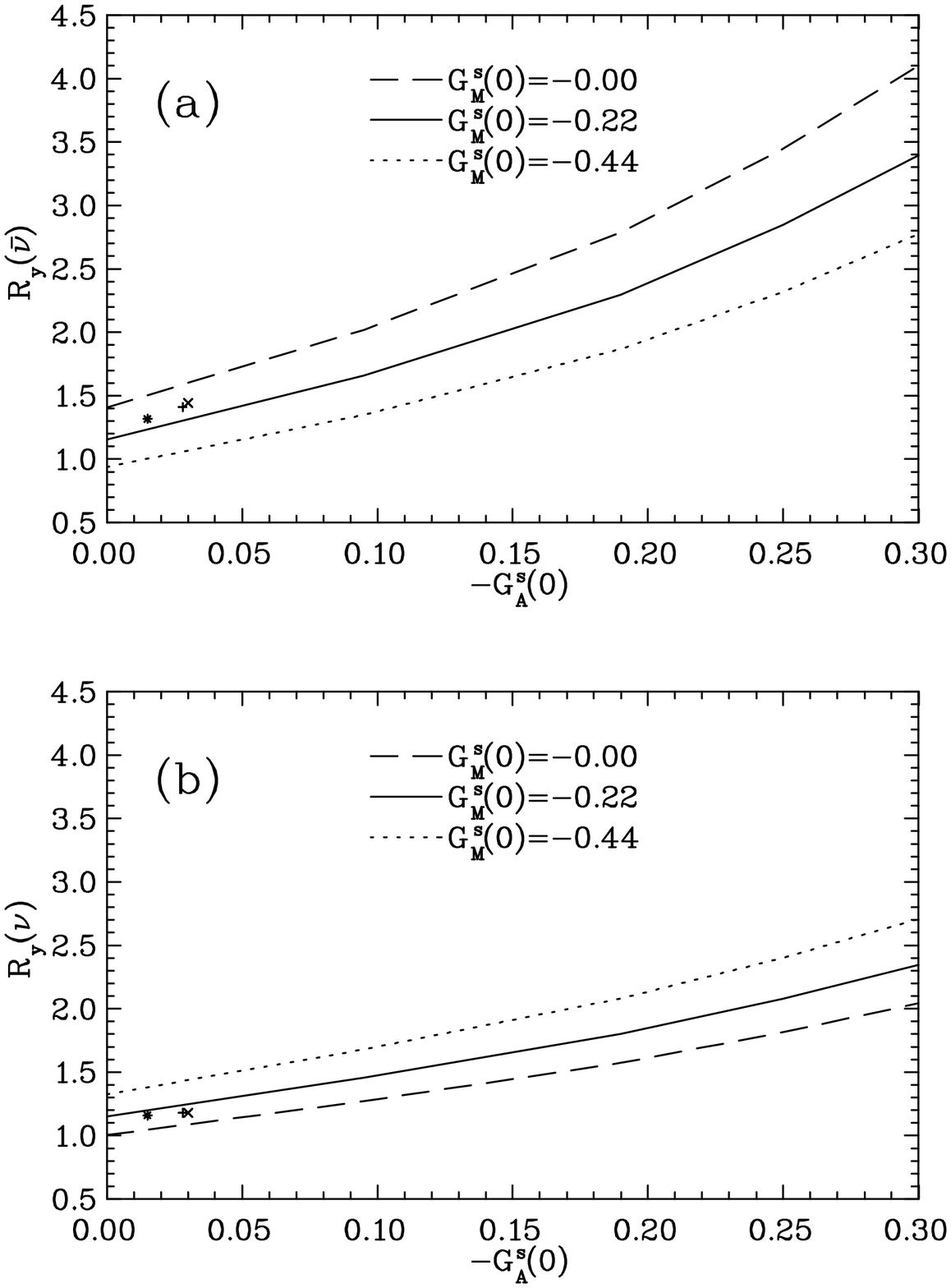,height=20.0cm,width=16.0cm}}
\end{figure}
\newpage
\begin{figure}[htb]
\centerline{\hskip -1.5cm
\epsfig{figure=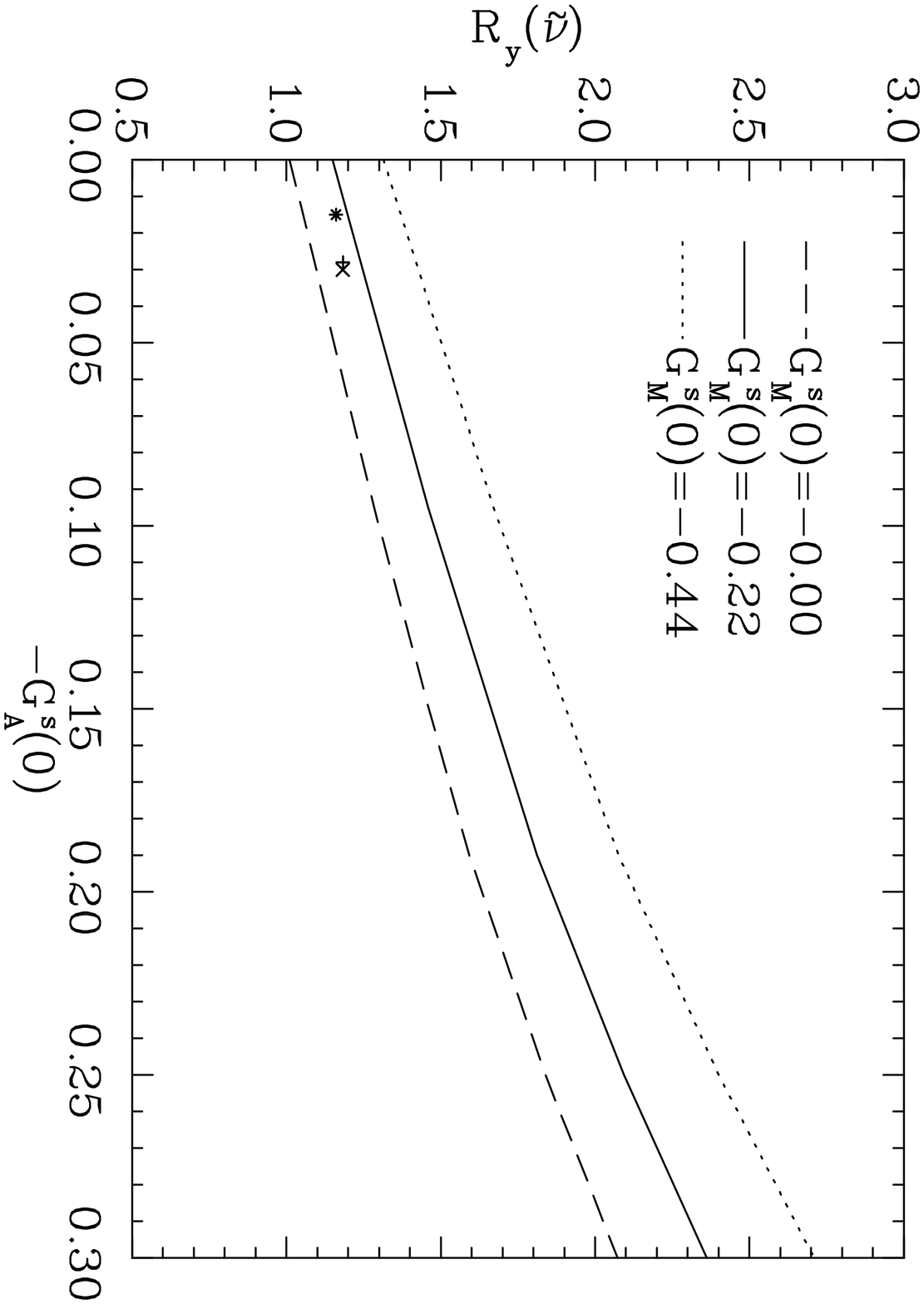,height=20.0cm,width=16.0cm}}
\end{figure}

\end{document}